\documentclass[nonacm, acmtog]{acmart}
\usepackage{graphicx} 
\usepackage[table]{xcolor}
\usepackage{booktabs} 

\title{Megakernel vs Wavefront GPU Path Tracing}
\author{Rafael Padilla}
\affiliation{\institution{University of Utah}\country{USA}}
\author{Kyle Webster}
\affiliation{\institution{University of Utah}\country{USA}}
\author{Austin H Kim}
\affiliation{\institution{University of Utah}\country{USA}}
\date{\today}

\begin{abstract}
    Over the last decade, advances in GPU hardware have been driven in large part by the demands of real-time graphics, culminating in dedicated hardware ray tracing cores (RT cores). These units accelerate ray-scene intersection queries directly in hardware, making physically based ray tracing algorithms increasingly practical for interactive applications. This paper compares and analyzes the performance of two ray-based rendering algorithms: forward path tracing (PT) and wavefront path tracing (WPT). GPU-based PT computes the color of each pixel by having each thread trace a single path to completion, naturally leading to a megakernel approach -- while WPT maintains state buffers between specialized kernel invocations to trace path stages simultaneously. We find that WPT affords a \textasciitilde16\% speedup over PT in our implementation. By analyzing traces from NVIDIA Nsight Graphics, we attributed this speedup to WPT's improved cache locality compared to PT. We also find that our implementation does not achieve maximum GPU throughput across any of its units, suggesting that communication and memory latency, as well as synchronization, are the limiting factors. Finally, we address potential algorithmic improvements and future work for real-time path tracing implementation for practical applications.
\end{abstract}

\begin{document}
\begin{teaserfigure}
\centering
\includegraphics[height=4cm]{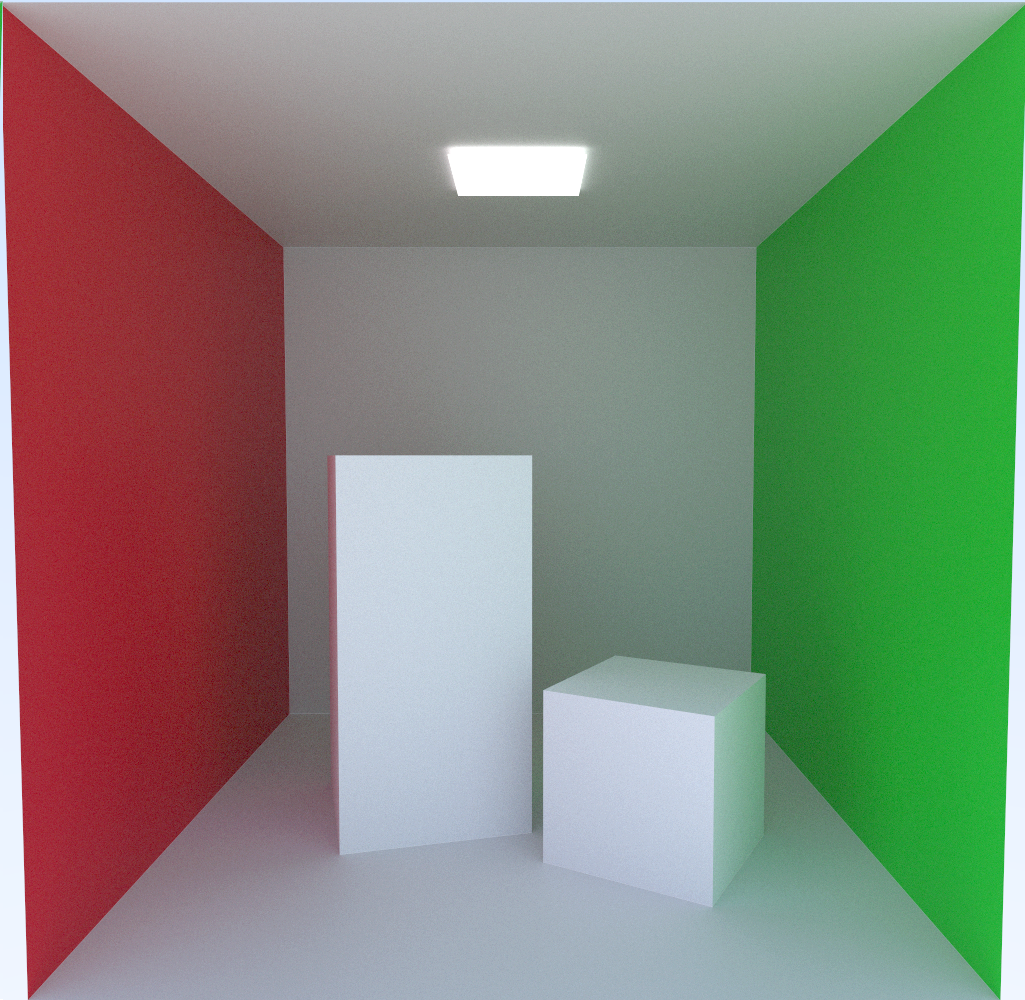}\hfill
\includegraphics[height=4cm]{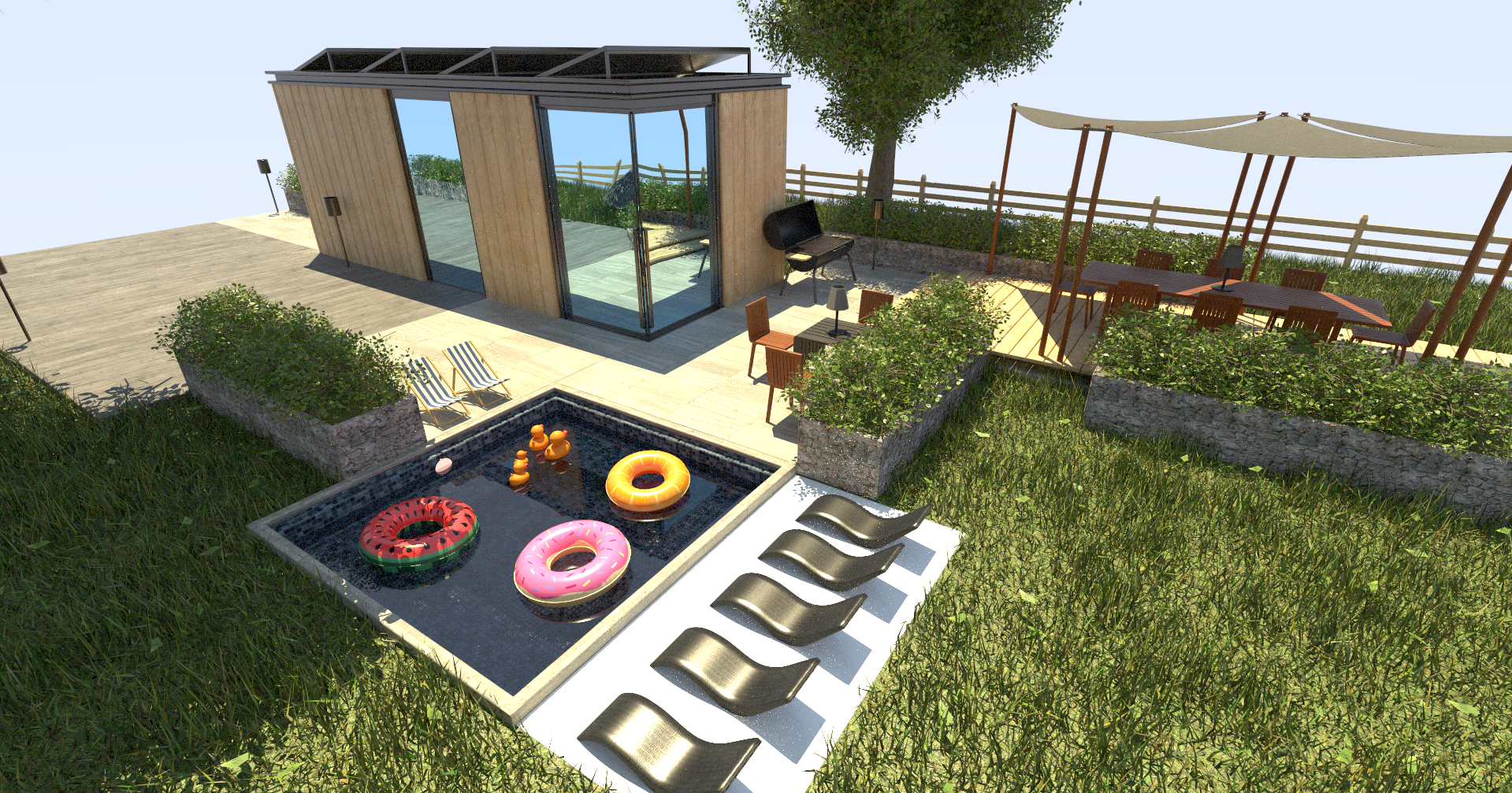}\hfill
\includegraphics[height=4cm]{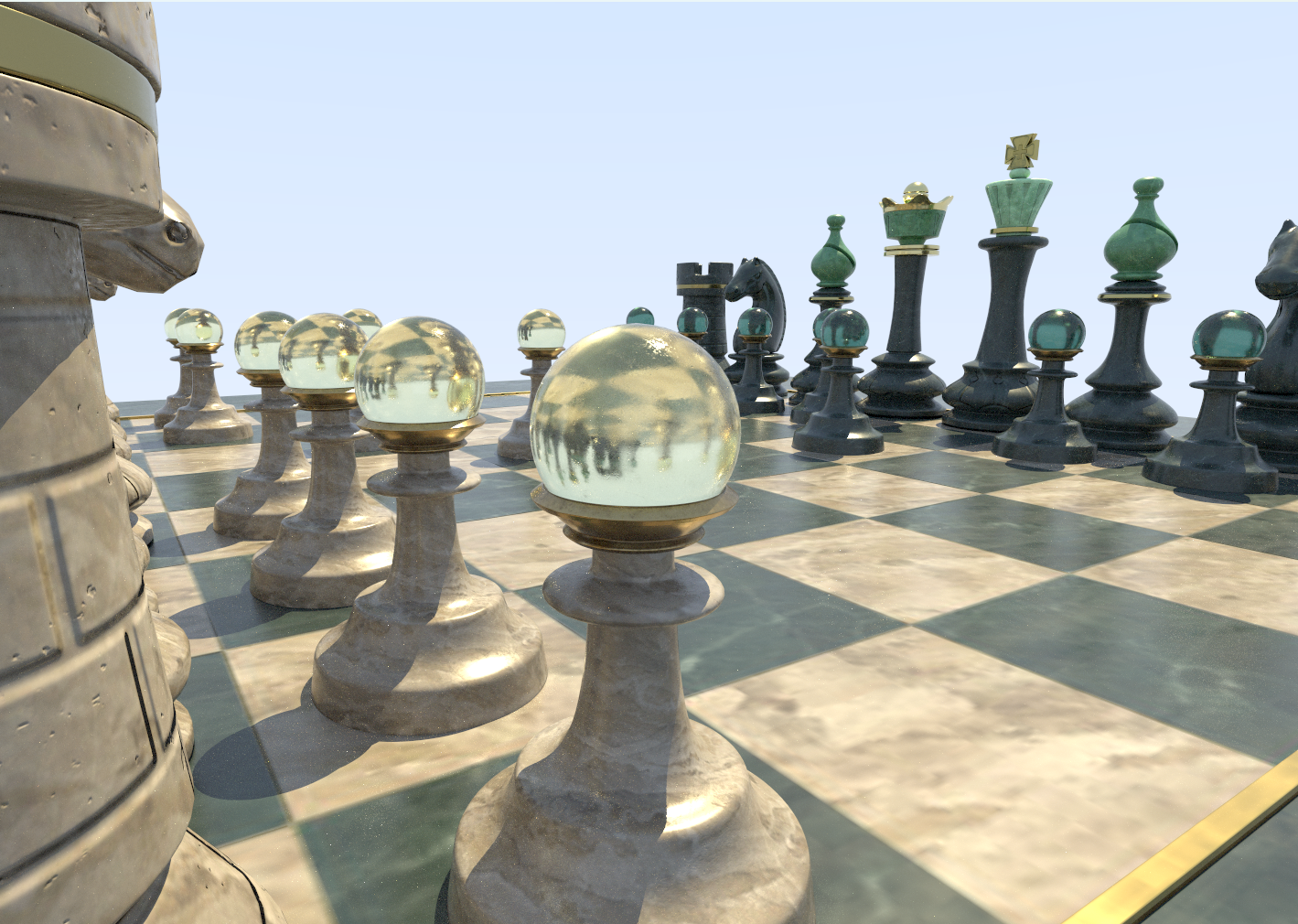}
\caption{Sample real-time generated images using our GPU hardware accelerated renderer.}
\label{fig:teaser}
\end{teaserfigure}
\maketitle

\section{Introduction}
Path tracing maps naturally onto massively parallel GPU hardware, yet efficient implementation remains challenging due to the irregular nature of light transport. As paths encounter different materials, bounce depths, visibility events, and termination conditions, threads within a warp can quickly diverge, reducing SIMD efficiency and often degrading memory locality. These issues become increasingly important on modern GPU architectures, where performance is frequently limited not only by arithmetic throughput, but also by latency hiding, cache behavior, and scheduling efficiency.

Wavefront path tracing \cite{megakernel} has emerged as an alternative formulation that addresses these issues by restructuring path tracing into coherent stages, grouping similar work into specialized kernels rather than tracing each path to completion independently. This approach has long been discussed in offline rendering and has become increasingly relevant for modern GPU-based rendering, yet its practical tradeoffs relative to simpler megakernel formulations remain dependent on architecture, implementation details, and workload characteristics. This motivates a direct study of how these approaches differ in practice.

In this paper, we implement both forward megakernel path tracing and wavefront path tracing within a shared Vulkan-based renderer and analyze their performance behavior on modern GPU hardware. Beyond comparing frame performance, we use NVIDIA Nsight Graphics to study hardware throughput, cache activity, and execution behavior to better understand the sources of observed differences. We further discuss how these tradeoffs relate to real-time versus offline rendering workloads, scene complexity, and broader considerations in GPU path tracing architecture design.

\section{Background}
There are many methods for synthesizing digital imagery, ranging from digital sketching and painting tools to light transport integrators on physically-based simulated scenes. Our renderer falls closer to the latter category. In this section we introduce the light transport equation, Monte Carlo integration, and the two algorithms we implement on the GPU that solve the rendering equation.

\subsection{Light Transport Equation}
Rendering can be formulated as a recursive integral equation that describes the transport of radiance between surfaces. That is, the \emph{Light Transport Equation} \cite{renderingeq}:
\begin{equation}
    L(x,\omega_o) = L_e(x,\omega_o)+\int_{S^2} L_i(x,\omega_i)f(x,\omega_o,\omega_i)|n\cdot\omega_i|\mathrm{d}\omega_i
\end{equation}
where $L$ is the radiance in direction $\omega_o$ at position $x$, $L_e$ is the emitted radiance, $L_i$ is the incident radiance, $f$ is a bidirectional scattering distribution function (BSDF), $n$ is the surface normal at point $x$ and the integral is over the sphere of directions $\omega_i$ at point $x$. This formulation of light transport lends itself to being solved by \emph{Monte Carlo Integrators}. A Monte Carlo Integrator provides an estimate of an integral by drawing samples from a distribution and evaluating the integrated function on those samples. The general form can be expressed as
\begin{equation}
    \hat{I} = \frac{1}{N}\sum_{i=1}^N \frac{f(X_i)}{p(X_i)}\approx\int_\Omega f(x)\mathrm{d}x
\end{equation}
Such an integrator is \emph{unbiased} if the support of $p(x)$ covers the support of $f(x)$ -- that is, the estimate will converge to the analytic value given enough samples if the generates samples can take any value which produce nonzero $f$. Both of the rendering algorithms that we implement are formulated as Monte Carlo Integrators.

\subsection{Forward Path Tracing}
The forward path tracing algorithm incrementally generates and evaluates paths from an initial camera ray -- accumulating the radiance traveling along that path to the camera and updating the \emph{throughput} from scattering at each surface. The Monte Carlo integrator appears as
\begin{equation}
    \hat{L}(\mathbf{x}_n) = \frac{L_{0\to1}T(\mathbf{x}_n)}{p(\mathbf{x}_n)},\quad T(\mathbf{x}_n) = \prod_{j=1}^{N-1} f(x_j,\omega_{o,j},\omega_{i,j})|n_j\cdot\omega_{i,j}|
\end{equation}
where $\mathbf{x}_n$ is a path of $n+1$ vertices $\{x_0, \cdots,x_n\}$ with $x_0$ on a light source, $p(\mathbf{x}_n)$ is the probability of generating path $\mathbf{x}_n$, and $T(\mathbf{x}_n)$ is the path throughput \cite{pbrt}.

The steps for performing forward path tracing are as follows:
\begin{enumerate}
    \item Generate initial ray from camera
    \item Extend path by intersecting scene in current direction $\omega$ to find scattering point $x$
    \item Accumulate radiance along path
    \item Sample new direction $\omega$ to extend path with, and update path throughput
    \item Repeat steps 2 through 4 until maximum path length is reached
\end{enumerate}
The megakernel implementation does the above steps from a single kernel dispatch, running one thread per pixel. In contrast, the wavefront implementation takes a batched approach.

\subsection{Wavefront Path Tracing}
Wavefront Path Tracing (WPT) decouples the path tracing pipeline into four sequential GPU compute stages: ray generation, scene intersection, material shading, and radiance accumulation. These are dispatched from the host application per bounce iteration. 

The ray generation stage initializes per-pixel ray state and samples primary ray directions through the camera projection model. Ray origins within each pixel are jittered using a Halton low-discrepancy sequence with bases 2 and 3, augmented with per-pixel Cranley-Patterson rotation to decorrelate spatially adjacent pixels, producing a well-distributed sample sequence across both pixels and frames that reduces visible noise at low sample counts compared to pseudorandom jitter.

The scene intersection stage traces each active ray against the scene's top-level acceleration structure (TLAS) using Vulkan inline ray queries, writing hit records to persistent per-pixel global buffers. The material shading stage then reconstructs full surface geometry from these stored records, evaluating interpolated normals, tangents, and texture coordinates without re-tracing the ray and evaluates the OpenPBR layered BSDF. Finally, the accumulation stage blends each completed path sample into a persistent buffer via a progressive running average, followed by tonemapping for display output. Between all stages, explicit Vulkan pipeline barriers ensure that buffer writes produced by each kernel are fully visible to the next before execution proceeds.

\section{Implementation}
Our renderer is implemented as a Vulkan-based GPU path tracer using compute shaders and hardware ray tracing via inline ray queries. Scene data, including geometry, transforms, materials, and acceleration structure handles, are packed on the CPU into GPU-friendly buffers and uploaded to the device. We use BLAS/TLAS acceleration structures to support efficient ray-scene intersection, while shading and light transport are evaluated in Slang shaders using a physically based path tracing integrator. Material response is modeled using Adobe OpenPBR, and assets are imported through Assimp, with scenes authored and tested in Blender. We implemented and compared two integrators within the same framework. In the forward path tracer, each GPU thread traces a path to completion in a megakernel-style design, performing intersection, shading, and path extension within a single kernel. In the wavefront path tracer, path states are instead stored in queues and processed across multiple specialized kernels corresponding to different path stages (e.g., ray generation, intersection, shading, and shadow evaluation). This enables more coherent execution and serves as the basis for our performance comparison. The system includes supporting infrastructure for experimentation and analysis, including a modular CMake-based build system, Vulkan resource management, and ImGui-based debugging interfaces.

\subsection{Wavefront Optimization}

An inefficiency that becomes apparent with wavefront path tracing is that all bounce iterations dispatch over the full pixel resolution, even though an increasing fraction of the paths terminate at each bounce due to Russian Roulette and scene misses. Dispatching W x H number of threads when only a small fraction perform useful work wastes GPU resources and increases memory traffic. 

This introduces two additional kernel stages per bounce, the compaction kernel and the indirect argument preparation kernel, but eliminates the wasted thread overhead that grows with bounce depth. For scenes with aggressive Russian Roulette termination, active ray counts drop substantially after the second or third bounce, producing net performance gains that increase with maximum bounce depth.

The compaction scheme employs a back and forth strategy: the active count buffer maintains two values, where index 0 holds the current bounce's active count and index 1 accumulates the next bounce's count during compaction. After compaction, the prepare-indirect kernel promotes index 1 to index 0 and resets index 1 to zero in preparation for the subsequent pass. The active index buffer is safely reused in-place, as each compaction thread reads its assigned index before any thread writes to a slot it might subsequently read, preventing race conditions without additional synchronization.

\begin{figure*}[h]
\centering
\begin{minipage}[b]{0.845\linewidth}
\includegraphics[width=\linewidth]{figures/tp\_mega.png}\\
\includegraphics[width=\linewidth]{figures/tp\_wave.png}
\end{minipage}
\includegraphics[width=0.15\linewidth]{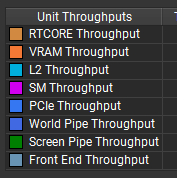}
\caption{{\bf Top:} Megakernel Alg. Nsight Graphics trace. {\bf Bottom:} Wavefront Alg. Nsight Graphics trace.}
\label{fig:placeholder}
\end{figure*}

\section{Performance Analysis}
Performance was evaluated using NVIDIA Nsight Graphics on an RTX 3060 Ti using the \textit{A Beautiful Game} scene from the Khronos glTF sample assets. Both the megakernel path tracer and wavefront path tracer were tested under identical scene and hardware conditions. Frame timing measurements show that the wavefront implementation achieved an average of 73.6 FPS (13.58 ms/frame), compared to 64.7 FPS (15.47 ms/frame) for the megakernel implementation, corresponding to an observed speedup of approximately 16\%.

Nsight throughput traces indicate differing hardware utilization patterns between the two approaches. The megakernel exhibited slightly higher SM throughput (37.1\% vs.\ 34.1\%) and RTCore throughput (16.9\% vs.\ 11.7\%), while the wavefront implementation showed substantially higher VRAM throughput (41.4\% vs.\ 19.3\%) and L2 cache throughput (22.5\% vs.\ 15.8\%). This suggests a shift toward greater memory-system activity and improved cache usage in the wavefront design. Despite these differences, neither approach saturated any major GPU execution unit, indicating underutilized compute and traversal resources across both methods.

\begin{table}[b]
\centering
\rowcolors{2}{blue!6}{blue!14}
\begin{tabular}{lcc}
\rowcolor{blue!35}
\textbf{Metric} & \textbf{Wavefront PT} & \textbf{Megakernel PT} \\
Average FPS        & 73.6  & 64.7 \\
Frame Time (ms)    & 13.58 & 15.47 \\
Relative Speedup   & 1.16$\times$ & 1.00$\times$ \\
\end{tabular}
\caption{Frame performance comparison on \textit{A Beautiful Game}.}
\label{tab:perf}
\end{table}

\begin{table}[b]
\centering
\rowcolors{2}{orange!8}{orange!18}
\begin{tabular}{lcc}
\rowcolor{orange!40}
\textbf{Nsight Throughput Metric} & \textbf{Wavefront} & \textbf{Megakernel} \\
SM Throughput        & 34.1\% & 37.1\% \\
RTCore Throughput    & 11.7\% & 16.9\% \\
VRAM Throughput      & 41.4\% & 19.3\% \\
L2 Throughput        & 22.5\% & 15.8\% \\
PCIe Throughput      & 12.2\% & 9.9\% \\
\end{tabular}
\caption{Selected GPU throughput metrics from NVIDIA Nsight Graphics.}
\label{tab:throughput}
\end{table}

\section{Discussion}
Our results show that wavefront path tracing achieved an approximately 16\% performance improvement over the megakernel forward path tracer in our implementation, despite neither method fully saturating GPU execution resources. The throughput data suggest that this improvement is less attributable to increased raw compute utilization and more consistent with improved memory behavior. In particular, the wavefront integrator exhibited substantially higher VRAM and L2 throughput, supporting the hypothesis that queue-based decomposition improved coherence and cache locality by grouping similar path stages together. This is consistent with the expected strengths of wavefront methods, which reduce divergence compared to monolithic megakernel execution, especially as paths branch due to material and transport complexity. Conversely, the megakernel showed somewhat higher SM and RTCore throughput, reflecting its more tightly coupled execution style, though this did not translate into superior overall frame performance.

These observations align with the theoretical tradeoffs between the two approaches. Forward path tracing in a megakernel is conceptually simple and minimizes explicit synchronization and kernel launch overhead, making it attractive for smaller scenes, simpler shaders, or latency-sensitive implementations. However, as scene complexity grows and rays exhibit increasingly divergent behavior, monolithic execution can suffer from poorer warp efficiency and reduced locality. Wavefront methods, by decoupling traversal, shading, and shadow evaluation into specialized stages, can better exploit coherence at larger scales, though at the cost of queue management overhead and more complicated scheduling \cite{megakernel}.

These tradeoffs also relate to application domain. In real-time and interactive rendering, megakernel approaches have historically been attractive because they minimize per-frame orchestration overhead and can reduce latency, particularly for lower sample count regimes. However, modern GPU architectures and increasingly complex material/light transport workloads have made wavefront approaches more compelling even in interactive contexts, as reflected in our measurements. For offline rendering, where throughput dominates over latency, wavefront-style decomposition is often even more advantageous, particularly for large scenes and deep path depths. Production renderers frequently use wavefront or related queue-based formulations precisely because they scale better under heavy transport complexity.

It is also important to distinguish GPU and CPU-oriented formulations. On GPUs, wavefront path tracing primarily addresses SIMT divergence and memory coherence. On CPUs, where divergence manifests differently and task scheduling models are more flexible, "wavefront" often overlaps conceptually with breadth-first or packet-style traversal approaches that improve cache behavior and load balancing across cores. CPU offline renderers have historically used both recursive megakernel-like integrators and wavefront variants, with the latter often favored for very large production workloads \cite{pixar}. The relative advantage is therefore architecture dependent, and likely sensitive to scene size, material complexity, ray depth, and scheduling overheads. Our results suggest that even for a moderately complex benchmark scene, these effects are already observable.

\section{Conclusion}
In this work, we implemented and analyzed two GPU path tracing architectures---forward megakernel path tracing and wavefront path tracing---within a shared Vulkan-based renderer using hardware ray tracing. Using NVIDIA Nsight Graphics and controlled experiments on the \textit{A Beautiful Game} benchmark scene, we observed that the wavefront implementation achieved an approximately 16\% performance improvement over the megakernel approach. Our analysis suggests this gain is associated less with higher raw compute utilization than with improved memory behavior and cache locality arising from the wavefront formulation.

More broadly, our results reinforce the tradeoff between simplicity and scalability in path tracing architecture design. While megakernel approaches remain attractive for their compactness and lower orchestration overhead, wavefront approaches offer advantages as transport complexity and divergence increase. At the same time, our throughput measurements indicate that neither implementation fully utilized available GPU hardware resources, pointing to latency, synchronization, and communication costs as remaining bottlenecks.

\subsection{Future Work}
Future work includes improving queue scheduling and memory layouts in the wavefront implementation, exploring larger and more diverse scenes, and investigating hybrid approaches that combine the low-overhead benefits of forward path tracing with the coherence advantages of wavefront methods. More broadly, these findings motivate continued study of GPU path tracing architectures for both offline rendering and emerging real-time applications.

\subsection{Contributions}
\begin{itemize}
    \item \textit{Rafael Padilla}: Literature review; Vulkan renderer architecture and implementation; megakernel forward path tracer implementation in Slang; OpenPBR material evaluation integration; NVIDIA Nsight Graphics profiling and performance analysis.
    \item \textit{Kyle Webster}: Initial wavefront path tracing algorithm design and Slang implementation; Intel VTune profiling.
    \item \textit{Austin H Kim}: Extended the wavefront path tracing pipeline with Multi-sample per pixel support; GPU stream compaction via atomic index lists and indirect dispatch to reduce idle thread overhead across bounce iterations.
\end{itemize}

\bibliographystyle{acm}
\bibliography{citations.bib}

\end{document}